\documentclass[aps,pre,twocolumn,superscriptaddress,floatfix]{revtex4-1}
\usepackage{amsmath,amsfonts,amssymb}
\usepackage{enumerate}
\usepackage{epsfig}
\usepackage{bbding}
\usepackage{textcomp}
\bibstyle{apsrev.bib}
\begin{document}

\title{Social Complexity: can it be analyzed and modelled?}

\author{Kimmo Kaski}
\affiliation{Centre of Excellence in Computational Complex Systems Research,
Department of Biomedical Engineering and Computational Science,
Aalto University}

\date{\today}

\begin{abstract}
Over the past decade network theory has turned out to be a powerful methodology to investigate complex systems of various sorts. Through data analysis, modeling, and simulation quite an unparalleled insight into their structure, function, and response can be obtained. In human societies individuals are linked through social interactions, which today are increasingly mediated  
electronically by modern Information Communication Technology thus leaving "footprints" of human behaviour as digital 
records. For these datasets the network theory approach is a natural one as we have demonstrated by analysing the dataset of multi-million user mobile phone communication-logs. This social network turned out to be modular in structure showing communities where individuals are connected with stronger ties and between communities with weaker ties. Also the network topology and the weighted links for pairs of individuals turned out to be related.These empirical findings inspired us to take the next step in network theory, by developing a simple network model based on basic network sociology mechanisms to get friends in order to catch some salient features of mesoscopic community and macroscopic topology formation.  Our model turned out to produce many empirically observed features of large-scale social networks. Thus we believe that the network theory approach combining data analysis with modeling and simulation could open a new perspective for studying and even predicting various collective social phenomena such as information spreading, formation of societal structures, and evolutionary processes in them.
\end{abstract}

\maketitle

\section{Introduction}
\label{intro}
As many biological and social systems consist of a large number of interacting constituents and show 
complex emergent and self-organized properties in structure, function, and response, one might ask whether 
their complexity is in the number of key elements. The answer is "No" since for example from biology we 
have learned that humans have 46 chromosomes, while the potato we eat has 48 and the cotton 
we wear has 52 or in other terms a roundworm (C. elegans) has nearly 20 000 and a mustard family plant 
(Arabidobsis) of about 27000 while humans have 23000 protein coding genes. On the other hand from sociology 
we have learned that the world with nearly seven billion inhabitants is a small world after all, since every one 
is separated on average by six steps from the others. So rather than having it in number it is the connectivities and 
their nature that matter and thus these systems could be viewed  as some sort of communication systems with 
many non-identical elements linked with diverse interactions. Hence these systems could be envisaged as networks.

With this view the question "how these complex systems could be studied" has an immediate answer {\em Network theory}, 
which has contributed and keeps contributing significantly to our understanding of their structural properties and 
dynamical processes in them  \cite{Ref1,Ref2}. For social systems of humans the network theory approach was 
introduced by social scientists and they established the key concepts and a number of tools to study mainly their structural 
properties \cite{Ref3,Ref4}. In a broader perspective of network approach it is the view of sociology that {\em social life consists of the flow and exchange of norms, values, ideas, and other social and cultural resources channeled through a network} \cite{Ref5}. Moreover, these networks - often with very complex topological structures - serve as substrates for various emergent, self-organizing, and collective dynamical phenomena of diffusion, spreading and co-evolution processes of e.g. news and epidemics, opinion formation, language evolution, etc. The inherent complexity of these systems in terms of structure, function, 
and response calls for computational network theory involving correspondingly data analysis, modelling, and simulation. 

Until recently the studied empirical data sets of social systems remained rather limited since the basic 
sources of data were questionnaires, thus the focus had been on smaller scale properties of communities 
rather than larger scale properties of whole societies. However, the recent development in information-communication 
technology (ICT) has opened the possibility to collect much larger societal level data sets from Internet, emails, phone records, 
etc. \cite{Ref6,Ref7,Ref8,Ref9,Ref10,Ref11}. While the scope of information in these "digital footprint" records is narrow as 
compared to detailed questionnaires, their huge amount and objective quantifiability enable us to study the social systems in the ways not possible before, including the investigation of the structure and dynamics of entire populations \cite{Ref12}. In these studies one has learned through data analysis quite a bit about the broad distributions of network characteristics, the small world properties, the modular organization of the social network in question, and the relationship between the network topology 
and the intensity of the ties in the net. 

With this information of the system at hand the next obvious question arrises naturally, namely "what are the mechanisms 
involved in generating its observed structural properties or function". To answer this question one needs to take the next step in computational network theory, i.e. modelling, which based on the view or belief that it can illuminate sociological questions. 
For building such a model one needs to ponder two further questions: (i) how simple the model can be to be able to describe some of the salient features of the system, and (ii) how the model can be validated. For answering the former question one can take the 
so called Einstein view, i.e. "as simple as possible but not simpler", and for answering the latter question one should compare at least qualitatively the results of the model with those found through analysis in the real system. Only then one might attempt the 
final step in computational network theory, namely simulation by using the developed model to predict the response of the system 
to some external influence. 

\section{Empirical analysis}
\label{analysis}
Human social systems can naturally be viewed as networks, where the nodes correspond to individuals and links to social interactions between them. These networks are known to have {\em Small World} property first described by Stanley Milgram through his 'six degrees of separation' experiment in 1967. Apart from this the pure connectivity related issue of topology of the network the social interactions in them have strength, which in turn reflects back not only to its structure but also to its formation and the dynamic processes taking place in them.

In studying social network systems we can ask two basic questions: (i) how are they organized, and (ii) can they be modelled with simple models. Answering the first question has traditionally been based on the analysis of data from questionnaires typically among N = 10$^2$ - 10$^3$ individuals with a wide scope of social interactions but with limited resolution in quantifying
the strength of interaction between a pair of individuals. In addition there is the problem that the tie strength may be view differently by the individuals of a socially interacting pair, or in other words the individuals have e.g. different scales of friendship. 

Alternatively one can take the  approach made possible by todays ICT, in which data is obtained from electronic records of interactions between typically of the order of N = 10$^6$ individuals or more in quantifiable and accurate fashion through measurements. Although in this case the scope of social interactions is narrower than with questionnaires, the quantity of data even with a few attributes of the individual subjects is usually so huge that new computational methodologies and tools need to be developed for data handling, accuracy, and wider perspective. This makes the ICT-based dataset studies complementary to 
those of questionnaires based studies thus enabling more comprehensive insight to social system. 

We have recently demonstrated the success of this type of ICT-enabled approach by studying social interaction network constructed from a very large data set of mobile phone communication logs  \cite{Ref11}. We have considered the social interaction to exist between individuals $i$ and $j$, if within a time period of 18 weeks $i$ calls to $j$ and $j$ returns at least a call to $i$ or visa versa; 
in other words we require reciprocity for a social link to exist. The strength of the social tie is measured either by the aggregate amount of time spent in calls ($w^T_{ij}$) or the total number of calls ($w^N_{ij}$) between a pair of individuals during the18 weeks period. In Fig. 1 we show a sample of this kind of construction using time as tie strength measure for about 1000 individuals of the total of 5 million subscribers of service of a mobile phone operator. 

\begin{figure*}[ht!]
  \begin{center}
     \includegraphics[width=12.0cm,angle=0]{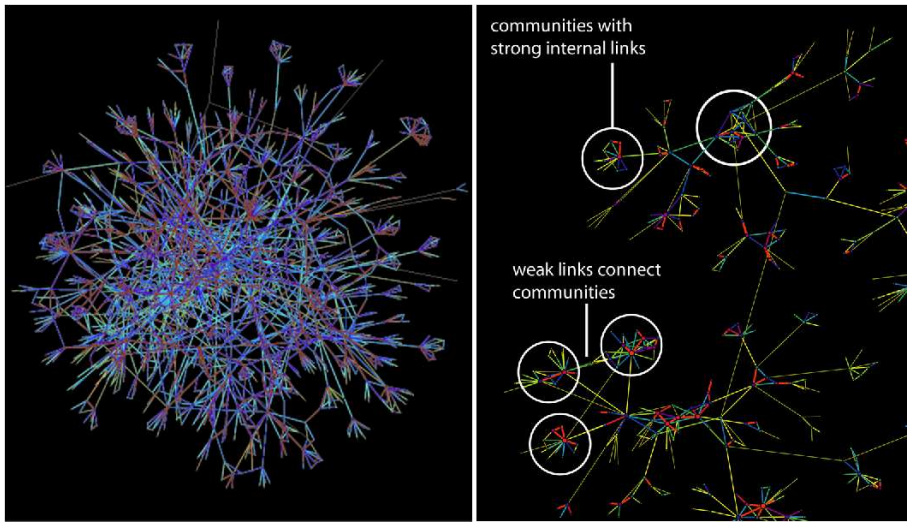}
  \end{center}
\caption{Left: A snowball sample of mobile phone dataset based social network with the strength
colour coded. Right: Enlargement of the sample indicating the structure of communities with stronger
internal links while the communities are connected with weaker links.}
\label{figure:1} 
\end{figure*}

In this figure (in the zoomed panel on the right) we see that apart from its global structure the network shows clear local structure 
due to tie strengths; the network shows quite high degree of modularity as communities with dense and stronger internal links are connected sparsely with weaker external links. It is thus evident that the weak and strong ties play different roles in a social network. Furthermore we have found that the network shows high degree clustering and assortative mixing, of which the latter demonstrates 
that high degree nodes in the network are connected to other high degree nodes or "popular people are highly connected to other popular people". 

Perhaps the most important finding of our empirical study is that we could verify that the network fulfils the strength of the weak ties hypothesis by Granovetter \cite{Ref3}, stating that {\em Tie strength between two individuals increases with the overlap of their friendship circles}. This makes it even more evident that the weak and strong ties have different roles, in fact to the extent as Granovetter further 
hypothesized that the weak ties maintain the global integrity of the network while strong ties maintain the communities.
Furthermore, we have demonstrated this difference in the roles of weak and strong ties by simple thresholding or percolation 
analysis where links were cut in either {\em descending} or {\em ascending} order of link weights. We found that when cutting links 
starting from stronger links and going towards weaker links, the largest connected component survives even up to 80 per cent of
all the links being cut. However, in contrast while going from the weaker links towards stronger links at 80 per cent of all links 
cut, only a number of small unconnected components remain \cite{Ref11}. 

In this study we also found the structural properties of the network to play an important role in its functional 
properties, e.g. in information spreading. We infected one node of the network with a piece of information 
and assumed its probability to hop randomly to a neighboring node to be dependent upon the strength 
of the link between them.  Here the link weight can be interpreted to correspond to the bandwidth of information 
transmission while in the reference case we assume all the strengths or link bandwidths to be constant. It turned out 
that in the reference case the information spread quickly through the whole system in the real system the information 
got trapped in local communities for long time making the spreading of information through the whole system much slower
over all.

\section{Modelling}
\label{sec:2}

As for the next step in computational network theory the above described empirical findings inspired us to develop a simple microscopic model to reproduce similar structural properties. Another reason for modelling research is that if the model turns out to be realistic in some ways it could then be used to simulate various dynamical processes in similar systems. The fact that the above empirical network is structured as modules or communities raises the question How do the communities emerge during the growth of the network. On the other hand the fact of its links being weighted raises the question how do they influence the formation of mesoscopic communities and macroscopic network topology. The empirical verification of Granovetters hypothesis links these two questions as how should we take into account the weight - topology correlation. 

In order to answer the above questions we have built a model, in which microscopic or individual-level friendship formation mechanisms translate on one hand to formation of mesoscopic communities and on the other hand to the whole macroscopic system. Although the actual processes taking place in forming social relations are undoubtedly very complex, the network sociology comes to help by identifying two fundamental mechanisms for network tie formation leading to network evolution: (i) cyclic closure forming ties at short range with ones network neighbours, and (ii) focal closure forming ties independently of the range through shared activities \cite{Ref13}. We adopted a simple scenario that new ties are created preferably through strong ties with every interaction making them even stronger, mimicking people getting acquaintances with local and global search mechanisms dependent on link weights.

In sociology the effect of link weight is complicated but it is reasonable to assume that people interact mostly through their strong friendships such that each interaction makes this connection even stronger. In reality it is also possible that some friendships disappear corresponding to deletion of a link between two individuals. However, this process is often so slow that it is difficult to observe from available data due to it being restricted to rather narrow time windows. In our model the number of links is reduced by deleting a network node and all the links connected to it with an adjustable probability corresponding to a slow rate of deletions. This could be viewed to correspond to a situation, in which the individual having belonged to the network for a while goes out of its scope (e.g. in case of mobile communication changing the carrier).

\begin{figure*}[ht!]
  \begin{center}
     \includegraphics[width=12.0cm,angle=0]{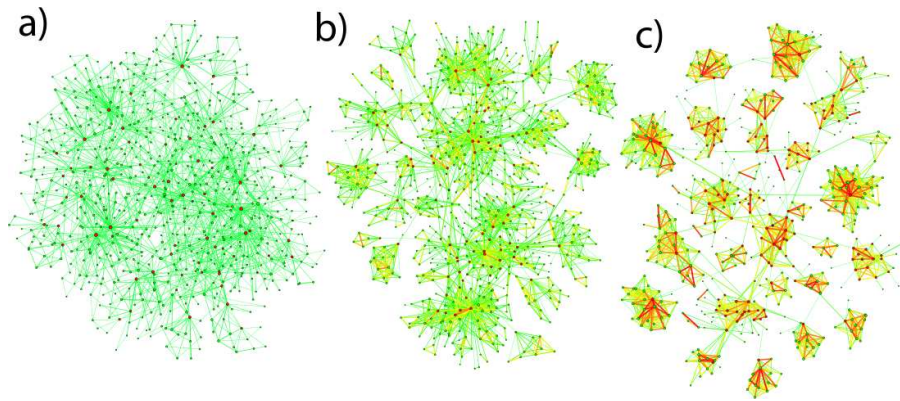}
  \end{center}
\caption{Visualization of the effect of friendship reinforcement on network topology. (a) no reinforcement,
(b) small reinforcement, and (c) large reinforcement (the average degree in these networks is 10). Weak links are green and 
the color changes gradually to red for strong links. (Published earlier as Fig. 4 in Comp. Phys. Comm. 180, 517 (2009); Copyright (2008) by Elsevier B.V., all rights reserved.)}
\label{fig:2} 
\end{figure*}

In our model \cite{Ref14} the weights of the existing links enter the model dynamics through 
reinforcement of visited links, which is controlled by the friendship reinforcement parameter. 
When it is large some links start to dominate the local search process
by attracting almost all searches and becoming all the time stronger. In this
case the search ends up almost always to a familiar node, that is, the start
and end nodes are already connected. Hence there is a tendency for the forming network to get 
locally structured as modules or communities. On the other hand, when the friendship 
reinforcement parameter is zero the local search follows all links with equal probability. 
Now the search exits the neighborhood of the start node quite easily, and a new link is then 
established with a certain probability. Thus there would be no tendency for local structuring. 

In Fig. 2 we depict the results of the final configurations of our network model when the friendship
reinforcement parameter is varied while all the other parameters of the model were kept fixed \cite{Ref15}. It is
clearly seen that increasing the reinforcement parameter promotes the formation of communities. 
Moreover, it turned out that strong links seem to be confined in communities while the links between
communities are mainly weak. In addition, the comparison between the final network topology configuration 
for large friendship reinforcement parameter value (Fig. 2c) and that of the mobile phone based social 
network turns our very favourably. Furthermore, in the model network we investigated all the same
properties as in the real mobile phone based social network and the comparison turned out very favorably 
by yielding similar results for all the measures used in both cases. So we were able to conclude that 
the microscopic mechanisms of our simple model serve as plausible explanation for the formation of
communities and other large scale structures of social systems including their emergent properties.     

\section{Concluding remarks}
\label{sec:L}
In summary we hope to have demonstrated that computational network theory is a powerful methodology 
to investigate various complex systems. It is our firm belief that the network theory through data analysis, 
modelling, and simulation can give quite an unparalleled insight into the structure, function, and response 
of these systems. To demonstrate this we presented some of our results on social systems, where we had  
analysed a huge dataset of mobile phone usage with the assumption that the system can be considered 
as a network of individuals interacting socially with measurable strengths. This empirical analysis showed 
the network to be structured as communities with strong internal ties and weak external ties between individuals 
of the network.  As further demonstration of the power of computational network theory we had developed a 
simple network model with basic sociology mechanisms to get friends included, producing many empirically 
observed features of the analyzed social networks. Hence believe that the computational network theory 
approach combining data analysis with modeling and simulation open a new and versatile perspective for 
studying and even predicting various collective social phenomena such as information spreading, formation 
of societal structures, and evolutionary processes in them.

\acknowledgements
This work has been funded by the Academy of Finland, under the Finnish Center of Excellence program 2006-2011,
proj. 129670.


\begin{thebibliography}{}
\bibitem{Ref1} M. E. J. Newman, A.-L. Barabasi, and D. J. Watts, The structure and dynamics of networks 
(Princeton University Press, Princeton, U. S., 2006).
\bibitem{Ref2} G. Caldarelli, Scale-Free Networks: Complex Webs in Nature and Technology (Oxford University Press, 
Oxford, U. K., 2007).
\bibitem{Ref3} M. Granovetter, Am. J. Sociol. 91, 481 (1985).
\bibitem{Ref4} S. Wasserman and K. Faust, Social Network Analysis: Methods and Applications (Cambridge University Press, Cambridge, U. K., 1994).
\bibitem{Ref5} H. White, S. Boorman, and R. Breiger, Am. J. Sociol. 81, 730 (1976).
\bibitem{Ref6} D. J. Watts and S. H. Strogatz, Nature (London) 393, 440 (1998).
\bibitem{Ref7} R. Albert, H. Jeong, and A.-L. Barabasi, Nature (London) 401, 130 (1999).
\bibitem{Ref8} H. Ebel, L.-I. Mielsch, and S. Bornholdt, Phys. Rev. E66, 035103(R) (2002).
\bibitem{Ref9} J.-P. Eckmann, E. Moses, and D. Sergi, Proc. Natl. Acad. Sci. U. S. A. 101, 14333 (2004). 
\bibitem{Ref10} P. S. Dodds, R. Muhamad, and D. J.Watts, Science 301, 827 (2003).
\bibitem{Ref11} J.-P. Onnela, J. Saramaki, J. Hyvonen, G. Szabo, D. Lazer, K. Kaski, J. Kertesz, and A.-L. Barabasi, Proc. Natl. Acad. Sci. U. S. A. 104, 7332 (2007).
\bibitem{Ref12} J. Bohannon, Science 314, 914 (2006); D. Butler, Nature (London) 449, 644 (2007).
\bibitem{Ref13} G. Kossinets and D. J. Watts, Science 311, 88, (2006).
\bibitem{Ref14} J.M. Kumpula, J.-P. Onnela, J. Saramaki, K. Kaski, and J. Kertesz, Phys. Rev. Lett. 99, 228701 (2007).
\bibitem{Ref15} J.M. Kumpula, J.-P. Onnela, J. Saramaki, J. Kertesz, and K. Kaski, Computer Physics Communications 180, 517 (2009). 

\end{thebibliography}
\end{document}